\documentclass{svjour3}                     
\smartqed  
\usepackage{graphicx}
\usepackage{mathptmx}      
\usepackage{amssymb}
\usepackage{amstext}
\usepackage{amsmath}
\usepackage{amsbsy}
\usepackage{makecell}
\usepackage{citesort}
\usepackage{color}
\usepackage[T1]{fontenc}
\newcolumntype{d}{D{.}{.}{-1}}

\newcommand*{\p}{\partial}
\newcommand{\mb}{\bar{\mu}_{\alpha \beta}^{nn'}}
\newcommand{\hh}{\frac{\hbar^2}{2}}
\newcommand{\eref}[1]{Eq.~(\ref{#1})}

\begin{document}
\title{Rovibrational interactions in linear triatomic molecules:\\
a theoretical study in curvilinear vibrational
coordinates\thanks{Dedicated to Professor Ma{\l}gorzata Witko on the occasion
of her 60th birthday}}

\titlerunning{Rovibrational interactions for triatomics}

\author{Leonid~Shirkov \and Tatiana~Korona \and Robert~Moszynski}

\institute{Leonid Shirkov \at
              Faculty of Chemistry, University of Warsaw,
Pasteura 1, 02-093 Warsaw, Poland,
              \email{leonid.shirkov@tiger.chem.uw.edu.pl}
\and
Tatiana Korona \at
              Faculty of Chemistry, University of Warsaw,
Pasteura 1, 02-093 Warsaw, Poland,
              \email{\mbox{tatiana.korona@tiger.chem.uw.edu.pl}}
\and
Robert Moszynski \at
              Faculty of Chemistry, University of Warsaw,
Pasteura 1, 02-093 Warsaw, Poland,
              \email{robert.moszynski@tiger.chem.uw.edu.pl} \\
}


\maketitle

\begin{abstract}
A variational solution to the rovibrational problem in curvilinear vibrational coordinates
has been implemented and used to investigate the nuclear motions in several
linear triatomic molecules, like HCN, OCS, and HCP.
The dependence of the rovibrational energy levels on the rotational quantum numbers and
the $l$-doubling has been studied.
Two approximations to the rovibrational
Hamiltonian have been examined, depending on the level of truncation of the potential
energy operator.
It turns out that truncation after the fifth order in the potential
is sufficient to produce vibrational energies of high accuracy.
An interesting feature of the present
formulation of the problem in terms of the curvilinear
vibrational coordinates is the explanation for the $l$-doubling
of the rovibrational levels, which in this picture is interpreted as the result of the inequivalency
of the average rotational constants in mutually perpendicular planes, rather than as the effect of the
Coriolis-type interactions between the vibrational and rotational motions.
The present theoretical results are compared with the available experimental data from
high-resolution spectroscopy, as well as with other {\em ab initio} calculations.

\end{abstract}

\section{Introduction\label{s1}}
A proper description of the interactions between the
rotational and vibrational motions is an important issue in molecular spectroscopy and the first attempts to find a theoretical solution date back to the beginning of the previous century.
Even a short overview of the most important methods is beyond the capacities of a single research paper,
so we therefore refer the reader to a recent tutorial overview of the existing approaches
\cite{tennyson2010}, and restrict ourselves to a necessarily incomplete
review of the literature
referring to successful implementations of the methods
directly related to the topic of our paper.

The choice of the variational method has several advantages over the perturbational theory
alternative.
Although in many cases   perturbation theory allows for a physically sound
interpretation of the results, as is the case with the rovibrational energy levels,
its possible drawbacks involve
among other things the convergence problems of the perturbation series, which do not appear
if the variational alternative is used.
The variational method is therefore much more robust, especially if many states are to be treated at once
and if Fermi resonances are expected.
One of the earliest attempts to solve the vibrational problem with the variational method was reported in Ref. \cite{carney1975} utilizing the Watson Hamiltonian \cite{wf1}.
Still other applications of this type can be found, e.g., in Refs. \cite{csaszar2007,csaszar2009}.
A solution to the full {\em rovibrational} problem for a linear molecule based on the isomorphic Watson Hamiltonian in the linear normal coordinates
with a basis set spanned by the 2$D$ harmonic oscillator eigenfunctions in a polar form
is in being sought in our laboratory \cite{Shirkov:unp}.
In Ref. \cite{carter1983} a method to solve
the variational rovibrational problem for linear triatomic molecules has been reported,
which was then developed in Refs. \cite{carter1990,martinez2006}.
However, the methods mentioned above  employ internal vibrational coordinates with the exact vibrational Hamiltonian, which make them difficult to extend to systems
with more than three atoms. The mixed variational-perturbative approach used by Suzuki \cite{sf1,sf2,sf3} should also be  mentioned in this context.

The well-known variational approach to triatomic molecules was implemented by Tennyson {\em et al.}
\cite{hf6,hf7}
within the {\sc dvr3d} software package,
which allows the solving of the full rovibrational problem. An example application of {\sc dvr3d} for
the  HCN and HNC molecules can be found in Ref. \cite{tennyson2002}.
However, in the method of Tennyson {\em et al.}
the Jacobi or Radau coordinates are used for the exact (within the Bohr-Oppenheimer approximation) Hamiltonian,
which makes it difficult to extend to molecules with a larger
number of atoms. It is also worth mentioning
the Hougen-Bunker-Johns approach for nonrigid molecules \cite{bunker1998,konarski1991}
implemented in the {\sc trove} package \cite{trove2007}. The illustrative application of this approach can be
found in Ref.~\cite{yurchenko2008} for HDO and Ref.~\cite{yachmenev2010} for HSOH.
However, in the Hougen-Bunker-Johns approach linear molecules should be treated
in a special way \cite{jensen1983}, and we are not aware of any implementation of this method
for this particular (linear) case.

The method presented in this paper is more universal
and can be applied to larger molecules, both linear and nonlinear.
It was proposed for the first time by Pavlyuchko \cite{pavlyuchko1988} and
is in principle limited only by the  computational costs
of the calculation of the potential energy surface (PES) and by the dimensions of the largest block
of the resulting rovibrational
Hamiltonian matrix. Some attempts to implement this method for linear molecules have been undertaken recently \cite{Pavlyuchko:conf}, but up to now no results obtained with this approach are available in the literature.
It is worth noting that many features of the rovibrational problem depend
on the way in which the vibrational and rotational coordinates are introduced.
The interpretation of the $l$-doubling is one of the prominent examples \cite{bf6,Watson:2001}.

The molecules selected for the first test of the new implementation of the rovibrational motion
problem are three triatomic molecules: HCN, HCP, and OCS.
The first two molecules contain a light hydrogen atom and show a large
amplitude vibration, while OCS is a rigid molecule with a low amplitude of the bending mode.
Additionally, for the HCN and HCP pair,
the effects of introducing a heavier atom belonging to the same group can be examined.

All three selected molecules are of considerable astrophysical and astrochemical interest.
Among them HCN is undoubtefully the most widespread one,
and numerous experimental and theoretical studies have been undertaken
regarding this molecule (see e.g., Refs. \cite{maki1974,Winnewisser:1976,maki1996,maki2000,mellau2008}).
It is also one of the most prominent interstellar and
circumstellar molecule. The degenerate bending
mode is an important feature of HCN because of the large amplitude of motion of the light
hydrogen atom and also because the bending motion is a direct pathway to the HNC isomer.

Theoretical and experimental data for the OCS molecule can be found
in Refs. \cite{maki1974,naim1998,Suvernev:1997,Muertz:2000}.
Spectroscopic interest in the OCS molecule stems from the fact that this molecule is present in the Venus atmosphere along with the more abundant carbon dioxide \cite{Krasnopolsky:2010}.
Experimental high-resolution rovibrational spectra have been recorded and assigned
e.g., in Refs. \cite{Belafhal:1995,hornberger1996,Rbaihi:1998,Strugariu:1998,Frech:1998,Toth:2010},
in some cases also for less common isotopomers of OCS.
One of the first theoretical analyses of the OCS spectra utilized the semiclassical approach
based on the integration of selected trajectories \cite{aubanel1988}.
Martin {\em et al.} \cite{Martin:1995} used the quartic force field obtained from the
coupled cluster calculations including single, double, and noniterative triple
excitations,
CCSD(T) \cite{Raghavachari:89}, with the cc-pVQZ basis set
to find the vibrational levels and some rotational constants for the OCS, CS, H$_2$S and CS$_2$ molecules by the second-order perturbation theory. They found for the carbonyl sulfide molecule
a generally good agreement with the available
experimental data \cite{Lahaye:1987}, although some Fermi resonances had to be accounted
for in order to get nearer to the experimental energy levels. For instance, the C=O stretching fundamental band ($\nu_1$) interacts with the $02^01$ combination band. Peterson {\em et al.} \cite{Peterson:1991} obtained a stretching PES for
the carbonyl sulfide molecule with the fourth-order M\o ller-Plesset perturbation theory (MP4) and
the size-consistency corrected configuration interaction method limited to single and
double excitations (CISD) and used them to
calculate the stretching bands of OCS as well as the rotational and
$l$-doubling constants.
Both stretching and bending modes have been examined by Pak and Woods
\cite{Pak:1997}, who used the CCSD(T) method to obtain the PES of OCS and the
second-order perturbation theory for the spectroscopic constants.
More recently, Xie {\em et al.} \cite{Xie:2001} devised a PES capable of
reproducing
high energy vibrations by
refining the force field constants of Pak and Woods \cite{Pak:1997} through the fitting
to experimental vibrational levels up to 8000~cm$^{-1}$.

The methylidynephosphine molecule HCP is an unstable species which decomposes after several hours
in room temperature conditions \cite{Gaumont:1994}, but it belongs to a small number of phosporus-containing
molecules present in the interstellar media \cite{Lattelais:2008} and is therefore of  considerable astrochemical interest.
From the chemist's point of view this
molecule is a rare example of
the existence of the unusual C$\equiv$P triple bond.
Interest in the spectroscopic characteristics of HCP grew in the early
nineties of the last century due to speculations about the presence
of this molecule in some planetary atmospheres \cite{Turner:1990}.
For its detection in such environments a detailed knowledge of the laboratory rotation-vibration spectrum is required.
Experimental examination of the HCP spectra
has been reported e.g. in Refs.  \ \cite{Tyler:1963,lehman1985a,jung1997,Drean:1996,Bizzocchi:2001}.
The quality of these measurements ranges
from the earliest low-resolution spectra to the most recent high-resolution ones.
The theoretical work on this molecule usually included the calculation
of the PES, which
in some cases was then utilized to reproduce vibrational levels and rotational constants
\cite{lehman1985a,koput1997,Drean:1996,Beck:2000}.
Koput and Carter \cite{koput1997} used the variational method for the vibrational problem followed by the
perturbation theory to describe the coupling between rotations and vibrations.
Specifically, perturbation theory
was applied to calculate the rotational constants with the PES calculated at the CCSD(T)
and the second-order M{\o}ller-Plesset perturbation theory (MP2) levels
of theory.
Koput and Carter \cite{koput1997} concluded that the MP2 level is not sufficient to reproduce the rovibrational spectra,
while the CCSD(T)
method is capable of doing so provided that large enough orbital basis is used in the
calculations.
The basis set effect is particularly important for the vibrational energy levels, while
the anharmonic constants are reproduced quite well in the smallest cc-pVTZ basis.
Koput and Carter \cite{koput1997} also tested the influence of the correlation contributions of the core electrons
to the vibrational and rotational parameters. They concluded that the core electrons' correlation
shifts the vibrational levels to higher energies and has little influence on other constants.
This line of study has been continued  by Beck {\em et al.} \cite{Beck:2000} who calculated the potential
energy surface with the
multireference configuration interaction method limited to single and double
excitations (MRCI) and used it to calculate
high vibrational states and the first rotational constant $B$.
Among other theoretical works on HCP one should mention
the study of polyads of highly excited vibrational states with the Fermi resonance Hamiltonian \cite{Joyex:1998}.

In this paper the implementation of the Rayleigh-Ritz variational solution for the rovibrational problem
with the Hamiltonian expressed in the curvilinear vibrational coordinates and
the Euler angles is reported and applied to a number of small triatomic molecules.
The simplicity of the proposed method follows from the polynomial representation
of the potential energy surface and the one-dimentional (1$D$) harmonic oscillator functions in the basis set.
In contrast to most of the other approaches employing the curvilinear coordinates, here the variational method is used
for the full rovibrational problem, and not just for its vibrational part, which allows for the larger
flexibility of the method compared to the perturbation theory.
With the present implementation
we calculated selected rovibrational levels of HCN, HCP, and OCS up to 13000~cm$^{-1}$.
For higher energies the polynomial representation of the potential energy operator
becomes inappropriate and one needs a Morse-like potential and Morse oscillator functions
in the basis set. In this case the matrix elements of the Hamiltonian become much more complicated
\cite{gribov1998,requena1986}.
However, we believe that the current implementation based on the polynomial expansion is sufficient to obtain
and interpret many low-energy rovibrational spectra.
In our investigation we paid a
special attention to the rovibrational interactions, caused by the dependence of the inverse inertia matrix on the vibrational coordinates.  The rovibrational levels were obtained up to and including $J=15$, then fitted to the known functional form including rotational centrifugal constants and constants of the $l$-resonance
and, finally, compared with the available experimental data.

The plan of this paper is as follows.
In section \ref{s2} we present the mathematical approach to solve the problem of nuclear motion
in polyatomic molecules based on the Hamiltonian written in the curvilinear vibrational coordinates
and show
some advantages
of this Hamiltonian over a Hamiltonian expressed in terms of linear vibrational coordinates. In section \ref{s3} we report the details
of the Rayleigh-Ritz variational procedure, discuss the choice of the basis set functions, and the way the potential energy function should be treated.
In section \ref{s4} the analytical fitting of the rovibrational energy levels,
which are widely used to represent the experimental data, are discussed.
The same analytical expressions are utilized to fit
the computed energy levels. Section \ref{s5} gives an explanation of the $l$-doubling effect
when the curvilinear coordinates are used. In section \ref{sec:tech}
we give some computational details of the potential energy calculations
and of the matrix diagonalization in the variational procedure. In section \ref{s6} we
compare our results with previous theoretical calculations and with the data
derived from high-resolution spectroscopic experiments.
Finally, section \ref{s7} concludes our paper.

\section{Molecular Hamiltonian\label{s2}}
Several ways of dealing with
the molecular Hamiltonian describing rovibrational motions have been described
in the literature, depending on whether
linear or curvilinear vibrational coordinates are used in the vibrational part.
In this section the theory in curvilinear coordinates will be presented.
We start the theory section by introducing the
curvilinear vibrational coordinates $Q_{s}$ \cite{McCoy:1991,gribov1998}
defined as linear combinations
of the natural (internal) vibrational coordinates $q_i$:
\begin{align}
&
q_{i}=\sum_{s}L_{is}Q_{s}, \quad \frac{\p}{\p q_{i}}=\sum_{s}(L^{p})_{is}\frac{\p}{\p Q_{s}}
\label{eq:qi}
\end{align}
The matrix $\mathbf{L}$ describes the mathematical form of the $s$-mode vibration
\cite{papousek1982,mills1972}.
The matrices $\mathbf{L}$ and $\mathbf{L}^{p}$ are related to one another by the following expression:
\begin{align}
&
(\mathbf{L}^{p})^{T}\mathbf{L}=\mathbf{I},
\label{eq:lp}
\end{align}
where $\mathbf{I}$ stands for the unit matrix.

The coordinates $Q_{s}$ are also known as {\em curvilinear} vibrational
coordinates, since they are related {\em nonlinearly} to the Cartesian coordinates of the atoms.
For linear molecules the natural vibrational coordinates $q_i$ are of two types,
corresponding to the bond stretching and to the linear angle bending modes.
The latter mode, which is doubly degenerate, is described by two coordinates in the mutually perpendicular planes \cite{mills1972}.
The kinetic energy operator for a polyatomic molecule can be written as
\cite{wilson1955,gribov1972}:
\begin{align}
&
\hat{T}=-\hh\sum_{i j}g^{1/4}\frac{\p}{\p q_{i}}g_{ij}(q)g^{-1/2}\frac{\p}{\p q_{j}}g^{1/4}.
\label{eq:hvib2}
\end{align}
The vibrational part of the Hamiltonian  can be reduced by performing
 the differentiation \cite{gribov1972,gribov1998}:
\begin{align}
&
\hat{H}_{v}=-\hh\sum_{i j}\frac{\p}{\p q_{i}}g_{ij}(q)\frac{\p}{\p q_{j}}-\frac{\hbar^{2}}{2}\beta(q)+V(q),
\label{eq:hvib3}
\end{align}
where $g_{ij}(q)$ denote elements of the kinetic matrix and
$\beta(q)$ is a nondifferential kinematic operator \cite{gribov1998}, also referred to
in the literature to
the pseudo-potential \cite{halonen1988}. The components of the $g$-tensor are well known
and can found in Ref. \cite{wilson1955}. Some features of $g$ for linear molecules are explained
in Refs. \cite{ferigle1951,gans1970}.

The exact dependence of the $g$-matrix on the vibrational coordinates is rather complicated,
therefore some approximations should be employed in \eref{eq:hvib3} in order to obtain tractable formulas.
Different approximate treatments of $\hat{H}_v$ in the curvilinear coordinates
have been proposed in the literature \cite{gribov1998}.
In this work we chose to expand the elements $g_{ij}(q)$ in the Taylor series:
\begin{align}
&
g_{ij}(q)=g_{ij}(0)+\sum_{k}\left(\frac{\p g_{ij}(q)}{\p q_{k}}\right)_{0}q_{k}+
\sum_{k l}\left(\frac{\p^2 g_{ij}(q)}{\p q_{k} q_{l}}\right)_{0}q_{k}q_{l}+\dots
.
\label{eq:taylor}
\end{align}
If
additionally the pseudopotential $\beta(q)$ is neglected,
the following expression for the vibrational Hamiltonian is obtained:
\begin{align}
&
\hat{H}_{v}=-\hh\sum_{s}\frac{\p^2}{\p Q_{s}^2}-\hh\sum_{s r t}\chi_{str}\frac{\p}{\p Q_{s}}Q_{t}
\frac{\p}{\p Q_{r}}-\frac{\hbar^2}{4}\sum_{s r t}\chi_{stur}\frac{\p}{\p Q_{s}}Q_{t}Q_{u}
\frac{\p}{\p Q_{r}}+V(Q)
,
\label{eq:hvib4}
\end{align}
where the kinematic coefficients $\chi_{spr}$ and $\chi_{sprt}$ in \eref{eq:hvib4} are defined by the formulas:,
\begin{align}
&
\chi_{spr}=\sum_{i j k}\left(\frac{\p g_{ij}(q)}{\p q_{k}}\right)_{0}L^{p}_{is}L^{p}_{jr}L_{kp}, \nonumber \\
&
\chi_{sprt}=\sum_{i j k l}\left(\frac{\p^2 g_{ij}(q)}{\p q_{k} \p q_{l}}\right)_{0}L^{p}_{is}L^{p}_{jt}L_{kp}L_{lr}
.
\label{eq:chi}
\end{align}

It should be noted that
for linear molecules the matrix of the $L_{kp}$ coefficients
is composed of two submatrices, corresponding to stretching and degenerate bending modes
with no cross-terms. This explains why the only non-vanishing components of $\chi$ are
$\chi_{sps}$ and $\chi_{spp's}$, the letters $s$ and $p$ refering to
the bending and stretching coordinates, respectively.

The total Hamiltonian describing the rovibrational motions of a polyatomic molecule can be expressed
through the curvilinear vibrational coordinates and the Euler angles $(\theta,\phi,\xi)$ which describe
the rotation of the axes of the equilibrium inertia tensor with respect to the laboratory frame
\cite{pavlyuchko1988}:
\begin{align}
&
\hat{H}=\hat{H}_{v}+\hh\sum_{\alpha\beta} \mu_{\alpha\beta}(Q)J_{\alpha} J_{\beta},
\label{eq:hrovib}
\end{align}
where $\mu_{\alpha\beta}(Q)$ denote elements of the inverse inertia tensor matrix.
To obtain a working expression for \eref{eq:hrovib}, we expand $\mu(Q)$ at the equilibrium point
in the curvilinear vibrational coordinates through the second order:
\begin{align}
&
\mu_{\alpha \beta}(Q)=\mu_{\alpha \beta}(0)+\sum_{r}\bigg(\frac{\p \mu_{\alpha\beta}(Q)}{\p Q_r}\bigg)_{0}Q_r+\frac12\sum_{r s}\bigg(\frac{\p^2 \mu_{\alpha \beta}(Q)}{\p Q_r \p Q_s}\bigg)_{0} Q_r Q_s + \dots
.
\label{eq:muab}
\end{align}
In order to find
the derivatives appearing in \eref{eq:muab} the expansion coefficients of the inertia tensor
in the linear normal coordinates $Q'_{r}$ are used:
\begin{align}
&
 I_{\alpha \beta}= I_{\alpha \beta}^0+\sum_{r}a_{r}^{(\alpha \beta)}Q'_{r}+
 \sum_{r s}A_{r s}^{(\alpha \beta)}Q'_{r}Q'_{s}+ \ldots
.
\label{eq:iab}
\end{align}
Explicit expressions for the coefficients
$a_{r}^{(\alpha \beta)}=\left(\frac{\p I_{\alpha \beta}}{\p Q'_{s}}\right)_{0}$
and $A_{rs}^{(\alpha \beta)}=\left(\frac{\p^2 I_{\alpha \beta}}{\p Q'_{r}\p Q'_{s}}\right)_{0}$
are well known in the literature \cite{tf8}.
The coefficients
$\bigg(\frac{\p \mu_{ab}(Q)}{\p Q_r}\bigg)_{0}$ and
$\bigg(\frac{\p^2 \mu_{\alpha \beta}(Q)}{\p Q_r \p Q_s}\bigg)_{0}$
can then be expressed through the $a_{r}^{(\alpha \beta)}$ and $A_{rs}^{(\alpha \beta)}$ terms,
cf. Ref.\ \cite{gribov1998} for the details.

We also have to note that linear molecules have $3N-5$ vibrational degrees of freedom and the
inertia tensor $I_{\alpha \beta}$ has only two main axes, $x$ and $y$. The vibrational modes can be classified as $N-2$
doubly-degenerate bending modes and $N-1$ stretching modes. Degenerate modes are described
by two coordinates in the perpendicular planes \cite{mills1972}. The internal bending coordinate is introduced as $\sin \theta_a$, where $\theta_a$ is the angle of the distortion
from linearity. In our {\em ab initio} calculations we used $\theta_a$ instead of $\sin \theta_a$,
because one can approximately set $\sin x \approx x$ for small values of $x$.

The rovibrational Hamiltonian expressed through the curvilinear vibrational coordinates,
Eq. (\ref{eq:hvib3}), does not contain Coriolis-type rovibrational interaction terms,
i.e. terms of the type
$J_{\alpha}p_{\beta}$ and $p_{\alpha}J_{\beta}$,
which are present in the rovibrational Watson Hamiltonian  derived in Refs. \cite{wf1,wf3}.
The Hamiltonian given by Eq. (\ref{eq:hrovib})
contains just the centrifugal interaction terms since $\mu_{\alpha \beta}(Q)$ depends on the vibrational coordinates only.
The other advantage of the latter Hamiltonian form is that there is no need for the special treatment
of linear molecules as in the case of the Watson Hamiltonian.

\section{The variational solution to the rovibrational problem\label{s3}}
In order to find the rovibrational energy levels
we have to solve the rovibrational problem described by the Hamiltonian given in \eref{eq:hrovib}, i.e. to solve the following Schr\"odinger equation:
\begin{align}
&
\hat H \Psi_{k}(Q,\phi,\theta,\xi) = {\cal E}_{k} \Psi_{k}(Q,\phi,\theta,\xi)
,
\label{eq:hpsiepsi}
\end{align}
where $k$ denotes a super-index containing quantum numbers corresponding to the vibrational ($n$) and
the rotational ($J,M,K,p$) degrees of freedom, i.e.\ $k=\{n,J,M,K,p\}$.
To solve the eigenproblem given by \eref{eq:hpsiepsi}
we follow the algorithm described in Ref.\ \cite{pavlyuchko1988}, where
the Rayleigh-Ritz variational method is applied. In this algorithm the wave function $\Psi_{k}$
is expanded in a basis of known functions $\varphi_{l}(Q,\phi,\theta,\xi)$:
\begin{align}
&
\Psi_{k}(Q,\phi,\theta,\xi)=\sum_{l}C_{kl}\varphi_{l}(Q,\phi,\theta,\xi)
,
\label{eq:psik}
\end{align}
where the basis function $\varphi_{l}(Q,\phi,\theta,\xi)$ is a product of a vibrational function $\psi_{n}(Q)$
and rotational function $\alpha_{MK}^{J}(\phi,\theta,\xi)$:
\begin{align}
&
\varphi_{l}(Q,\phi,\theta,\xi)=\psi_{n}(Q)\alpha_{MK}^{J}(\phi,\theta,\xi)
.
\label{eq:varphi}
\end{align}
The functions $\alpha_{MK}^{J}(\phi,\theta,\xi)$ are solutions to the Schr{\"o}dinger equation for the symmetric top
\begin{align}
&
\alpha_{MK}^{J}(\phi,\theta,\xi)=\sqrt{\frac{2J+1}{8\pi^2}}D_{MK}^{(J)^\star}(\phi,\theta,\xi),
\label{eq:alphajmk}
\end{align}
where $D_{MK}^{(J)}(\phi,\theta,\xi)$ are the Wigner $D$-functions \cite{zare1988}.
The dimension of the basis set $\{\varphi_l\}$ can be minimized if the vibrational functions
$\psi_{n}(Q)$ are variational solutions to the multidimensional vibrational Schr\"{o}dinger equation:
\begin{align}
&
\hat{H}_{v}\psi_{n}(Q)=E_{n}\psi_{n}(Q).
\label{eq:hv}
\end{align}
In \eref{eq:hv}
$n$ stands for several vibrational indices, e.g.\ for a linear triatomic molecule $n=\{n_{1},n_{2a},n_{2b},n_{3}\}$. Here $n_{2a},n_{2b}$ are the indices
of the basis set functions describing the motion in the perpendicular planes of the degenerate bending mode.

The vibrational problem, \eref{eq:hv}, is solved in a basis set constructed
from the products of the 1$D$ harmonic oscillator eigenfunctions $\phi_l(\tilde Q)$
for each vibrational degree of freedom:
\begin{align}
&
\phi_{n_s}(\tilde{Q}_{s})=N_{n_s}H_{n_s}(\tilde{Q}_{s})\exp[-\tilde{Q}_{s}^{2}/{2}], \quad \tilde{Q}_{s}\in]-\infty,\infty[
,
\label{eq:phil}
\end{align}
where $H_n(\tilde{Q}_{s})$ is the Hermite polynomial of the $n$th order and
$N_{n}=(2^{n}n!\sqrt{\pi})^{-1/2}$ is the normalization constant.
We have also introduced $\tilde{Q_r}$, a dimensionless
coordinate, $\tilde{Q_r}=Q_r\frac{k_r^{1/4}}{\hbar^{1/2}}$, $k_r$ is the harmonic
force constant for the $r$th vibrational mode, and $\hbar$ the Planck constant
in the appropriate unit system.
In the particular case of a linear triatomic molecule the multidimensional vibrational wave function is
expanded as:
\begin{align}
&
\psi_{k}(\tilde{Q}_1,\tilde{Q}_{2a},\tilde{Q}_{2b},\tilde{Q}_3)=
\sum_{n}C_{n}\phi_{n_1} \phi_{n_{2a}} \phi_{n_{2b}} \phi_{n_3}
.
\label{eq:psin}
\end{align}

The last remaining issue for the pure vibrational problem is the representation
of the potential energy operator $V$. Usually the Taylor expansion around the equilibrium point
truncated after the lowest few terms
is used in practical calculations.
The Taylor expansions of the potential energy function in both the internal and
curvilinear coordinates
are listed below:
\begin{align}
&
V(q)=\frac12\sum_{i j}f_{ij}q_{i}q_{j}+\frac16\sum_{i j k}f_{ijk}q_{i}q_{j}q_{k}+
\frac{1}{24}\sum_{i j k l}f_{ijkl}q_{i}q_{j}q_{k}q_{l}+\ldots
,
\label{eq:vq}
\end{align}
\begin{align}
&
V(Q)=\frac12\sum_{s p}F_{sp}Q_{s}Q_{p}+\frac16\sum_{s p r}F_{spr}Q_{i}Q_{j}Q_{k}+
\frac{1}{24}\sum_{s p r t}F_{sprt}Q_{s}Q_{p}Q_{r}Q_{t}+\ldots
.
\label{eq:vQ}
\end{align}
With the help of Eqs. (\ref{eq:vq}) and (\ref{eq:vQ}) the expansion coefficients
$f_{i\ldots}$ (for the internal coordinates)
and
$F_{i\ldots}$ (for the curvilinear vibrational coordinates) are implicitly defined.
The $F_{i\ldots}$ coefficients are obtained from the $f_{i\ldots}$ coefficients, which in turn are
calculated by an {\em ab initio} method.
The corresponding transformation formulas for the curvilinear vibrational coordinates are given
by the following equations \cite{mills1972}:
\begin{align}
& F_{sp}=\sum_{i j}f_{ij}L_{is}L_{jp}, \nonumber \\
& F_{spr}=\sum_{i j k}f_{ijk}L_{is}L_{jp}L_{jr}, \nonumber \\
& F_{sprt}=\sum_{i j k l}f_{ijkl}L_{is}L_{jp}L_{jr}L_{jt}
.
\label{eq:fnoprim}
\end{align}
This transformation is simpler than the analogous expression for the linear vibrational coordinates \cite{watson2001}
since no higher-order components of the $\mathbf{L}$ tensor appear in \eref{eq:fnoprim}.
Of course,
in the harmonic approximation both approaches (linear and curvilinear)
are fully equivalent.

In a finite basis of the functions $\varphi_{l}$,
the solution of the rovibrational problem, \eref{eq:hpsiepsi}, is reduced
to the diagonalization of the Hermitian matrix with the following matrix elements:
\begin{align}
&
H_{nK,n'K'}=\langle\psi_{n}(Q)\alpha_{M_JK}^{J}(\phi,\theta,\xi)|\hat{H}|\psi_{n'}(Q)\alpha_{M_JK'}^{J}(\phi,\theta,\xi)\rangle.
\label{eq:helem}
\end{align}
This general formula can greatly be simplified  if
$(i)$ the structure of the rovibrational Hamiltonian, \eref{eq:hrovib},
$(ii)$ the properties of the vibrational basis functions, and
$(iii)$ the known selection rules for the Wigner $D$-functions
are all taken into account.
After some algebra \cite{pavlyuchko1988},
the following equation for the element of the Hamiltonian matrix is obtained:
\begin{align}
&
H_{nK,n'K'}=E_{n}\delta_{nn'}\delta_{KK'}+\hh\sum_{\alpha\beta}\bar \mu_{\alpha,\beta}^{nn'}
\langle J,K|J_{\alpha}J_{\beta}|J,K'\rangle
,
\label{eq:helem2}
\end{align}
where the $\mb$ quantities
are defined as:
\begin{align}
&
\mb=\langle n|\mu_{\alpha\beta}(Q)|n'\rangle=\int \psi_{n}^\star(Q)\mu_{\alpha\beta}(Q) \psi_{n'}(Q)dQ
.
\label{eq:mb}
\end{align}
It should be stressed that
$\mb$ are
the coefficients that determine the rovibrational interactions.
The diagonal terms in \eref{eq:mb}
can be interpreted as
effective rotational constants for the $n$th vibrational state,
while the off-diagonal terms are responsible for the interactions of different vibrational energy levels with rotational motion.
All coefficients appearing in the equations above can be found in Ref. \cite{pavlyuchko1988}.

For $J=0$ the diagonalization of the Hamiltonian matrix gives
pure vibrational energy levels.
For states with $J>0$ rovibrational interactions appear.
Their physical nature is centrifugal since the elements of the matrix $\mu(Q)$ from \eref{eq:hrovib}
depend on the vibrational coordinates~$Q$.
Obviously,
neglecting of the dependence of the matrix $\mu(Q)$ on the vibrational coordinates by taking
only the first term  in the expansion given by \eref{eq:muab},
makes the matrix $\bar \mu$, \eref{eq:mb}, diagonal. In this case
the rovibrational problem can be separated into pure vibrational and
pure rotational problems.

\section{Constants of rovibrational interactions for linear triatomic molecules\label{s4}}
The formulas for the constants of rovibrational interactions and for  $l$-doubling are
well known in the literature. However, for the sake of completeness and to introduce
the notation we repeat them here.
The rovibrational levels for the linear top have been derived in Refs. \cite{amat1958_1,amat1958_2} and are taken in the present form from Refs. \cite{maki1996,maki2000,zelinger2003,mellau2008}.

The rotational energies for a given vibration, as obtained from  \eref{eq:hpsiepsi}, can be expressed in the following form:
\begin{align}
&
E_{r}=E_{r}^{\rm d}+E_{r}^{\rm n-d}
,
\label{eq:er}
\end{align}
where $E_{r}^{\rm d}$ and $E_{r}^{\rm n-d}$ denote diagonal and non-diagonal parts of the expression, respectively.
The diagonal part is given by the following formula:
\begin{align}
&
E_{r}^{\rm d}=B_{\nu}[J(J+1)-K^2]-D_{\nu}[J(J+1)-K^2]^{2}+H_{\nu}[J(J+1)-K^2]^{3}
,
\label{eq:erdiag}
\end{align}
where $\nu$ denotes the vibrational quantum numbers, $\nu=\{\nu_1,\nu_2^{\pm l},\nu_3\}$,
$B_{\nu}$ is the rotational constant, and $D_{\nu}$ and $H_{\nu}$
are the quadratic and cubic constants of the centrifugal distortion, respectively.
The non-diagonal term was fitted to the following expression \cite{mellau2008}:
\begin{align}
&
\langle \nu_1,\nu_2,\nu_3,l,J|\mathcal{H}| \nu_1,\nu_2,\nu_3,l\pm2,J\rangle=\pm\frac12[q_{\nu}-q_{\nu J}J(J+1)+q_{\nu JJ}J^2(J+1)^2+q_{l}(l\pm1)^2] \nonumber \\
&\times \big[(\nu_2\pm l)(\nu_2\pm l+2)(J(J+1)-l(l\pm 1))(J(J+1)-(l\pm 1)(l\pm 2))\big]^{1/2}
.
\label{eq:ernodiag}
\end{align}
Here, $\mathcal{H}$ stands for some effective perturbation Hamiltonian used in Ref. \cite{mellau2008}.
In the particular case of the (01$^{\pm1}$0) vibrational states we have:
\begin{align}
&
E_{r}^{\rm n-d}=\pm\frac12[q_{\nu}-q_{\nu J}J(J+1)+q_{\nu JJ}J^2(J+1)^2]J(J+1)
.
\label{eq:er0}
\end{align}
More formulas for the vibrational states (02$^{0}$0), (02$^{\pm 2}$0) and (0$\nu^{\pm 1}$0) can
be derived from \eref{eq:ernodiag}.
The nonzero constants $D_{\nu}$, $H_{\nu}$, $q_{\nu}$, $q_{\nu J}$, and $q_{\nu JJ}$
result from the dependence of $\mu_{\alpha \beta}$ on $Q$, cf. \eref{eq:muab}.
It should be noted that
the variational approach allows the calculating of the dependence of $D$ and $H$ on $J$,
while perturbation theory allows us to find these constants only for the ground vibrational state
\cite{carney1975,koput1997}.

\section{Symmetry considerations and $l$-doubling\label{s5}}
Triatomic linear molecules have three normal modes: two stretching vibrations usually
denoted by $\nu_1$ and $\nu_3$ and one degenerate bending vibration denoted by $\nu_2$.
Depending on the presence or absence of a center of symmetry, linear
molecules have either $D_{\infty h}$ or  $C_{\infty v}$ symmetry.
For noncentrosymmetric molecules
two stretching vibrations belong to the $\Sigma^{+}$ representation, while
the bending has  $\Pi$ symmetry \cite{bunker1998}. The stretching vibrations in this case
can usually be approximately attributed to a particular bond, e.g. for the HCN molecule
$\nu_1$ corresponds approximately to the C--H stretch and $\nu_3$ to the C$\equiv$N stretch.

A few words of explanation are due as far as
the nature of the $l$-doubling is concerned.
Commonly \cite{wf1,bf6},
when the rovibrational Watson Hamiltonian for linear molecules is used \cite{wf2},
this effect is described by the Coriolis interactions of the rotational
and vibrational motions of the molecule.
The explanation of the $l$-doubling when the Hamiltonian is expressed in curvilinear vibrational coordinates
is different since, as has already been pointed out in Section \ref{s2}, there is no Coriolis coupling in \eref{eq:hrovib}.
In the current approach this effect is explained
by the fact that for degenerate molecular deformations in two directions
the average values of the inertia tensor are different.
As a consequence, for a given vibrational state rotational states
with the same absolute value, but with a different orientation
of the projection of angular momentum on the symmetry axis $K$,
have different values of the rotational constants $\mb$ and therefore also different rotational energy levels.
These two possible explanations for the $l$-doubling have been
pointed out by
Herzberg \cite{bf6}.

\section{Computational details\label{sec:tech}}
Force constants corresponding to the
potential energy function $V(q)$, \eref{eq:vq}, were calculated at the CCSD(T) level
with the {\sc molpro} suite of programs \cite{MOLPRO2010}.
The cc-pVQZ basis set was used for all atoms.
Only valence electrons were correlated in the CCSD(T) calculations.
Since no analytical gradients are available in {\sc molpro} for the CCSD(T) method,
finite-difference formulas were applied in order to obtain the appropriate force constants.
This technique naturally limits the maximum order of the force constants that can be calculated with sufficient precision.
In order to check the accuracy of the numerical procedure, both 5- and 7-point formulas were
used for one-variable derivatives and multi-point formulas for mixed derivatives.
Then, only the converged digits were utilized in the subsequent calculations.

The formulas for some exemplary finite-difference fourth derivatives used in this work are
given below, where it is assumed that the potential energy function in the internal coordinates
is given by $V(r_1,r_2,\theta_{a})$.
In the derivative calculations
the distance increments $h, h_1$, and $h_2$ were set equal to 0.01~\AA \ and the angle increment $a$
to $1^{\circ}$.
\begin{align}
&
\frac{\partial^4 V}{\partial r_1^4}\Big|_{0} =
\Big(
6 V(0,0,0) -4 \big( V(h,0,0) + V(-h,0,0) \big)
+\big( V(2h,0,0) + V(-2h,0,0) \big)
\Big)/h^4
\nonumber \\
&
+ \mathcal{O}(h^6),
\label{eq:V41}
\end{align}
\begin{align}
&
\frac{\partial^4 V}{\partial r_1^4}\Big|_{0} =
\Big(
56 V(0,0,0) -39 \big( V(h,0,0) + V(-h,0,0) \big)
\nonumber\\
&
+12 \big( V(2h,0,0) + V(-2h,0,0) \big)
-\big( V(3h,0,0) + V(-3h,0,0) \big)
\Big)/(6h^4) + \mathcal{O}(h^8),
\label{eq:V42}
\end{align}
\begin{align}
&
\frac{\partial^4 V}{\partial r_1^2\partial r_2 \partial \theta_{a}}\Big|_{0} =
\Big(
    V( h_1, h_2, a)
   -V( h_1,-h_2, a)
   +V( h_1,-h_2,-a)
   +V(-h_1, h_2, a)
\nonumber \\
&
   -V(-h_1,-h_2, a)
   -V(-h_1, h_2,-a)
   +V(-h_1,-h_2,-a)
+2 \big(
   V(   0, h_2, a)
 - V(   0,-h_2, a)
\nonumber \\
&
 - V(   0, h_2,-a)
 + V(   0,-h_2,-a)
\big)
\Big)/(4h_1^2 h_2 a)
  + \mathcal{O}(h_1^2 h_2^2 a^2),
\label{eq:V43}
\end{align}
\begin{align}
&
\frac{\partial^4 V}{\partial r_1^2\partial r_2 \partial \theta_{a}}\Big|_{0} =
\Big(
16\big(
    V( h_1, h_2, a)
   -V( h_1,-h_2, a)
   +V( h_1,-h_2,-a)
   +V(-h_1, h_2, a)
\nonumber \\
&
   -V(-h_1,-h_2, a)
   -V(-h_1, h_2,-a)
   +V(-h_1,-h_2,-a)
\big)
+30\big(
 - V(   0, h_2, a)
\nonumber \\
&
 + V(   0,-h_2, a)
 + V(   0, h_2,-a)
 - V(   0,-h_2,-a)
\big)
   -V( 2h_1, h_2, a)
   +V( 2h_1,-h_2, a)
\nonumber \\
&
   +V( 2h_1, h_2, a)
   -V( 2h_1,-h_2,-a)
   -V(-2h_1, h_2,-a)
   +V(-2d_1,-h_2, a)
\nonumber \\
&
   +V(-2h_1, h_2,-a)
   -V(-2h_1,-h_2,-a)
\Big)/(48 h_1^2 h_2 a)+\mathcal{O}(h_1^4 h_2^2 a^2)
.
\label{eq:V44}
\end{align}

In order to estimate the accuracy of the Taylor expansion for $V$, the series
truncated after the fourth and the fifth orders have been used in the
production calculations. They are referred to as Approximation~1 and Approximation~2,
respectively. Application of Approximation~3, i.e. the series truncated after the
sixth order, is unfortunately not possible because of the limited accuracy of the numerical
differentiation.
Since the calculations reveal that
for molecules involving hydrogen atoms the contributions from the fifth order are significant,
and hence the sixth-order terms should possibly be included,
we expect somewhat larger errors in the rovibrational energy levels for the HCN and HCP molecules.

The rovibrational energies were calculated with a program written by one of our team
(L.S.). The masses of the most common isotopes were used in the calculations of the rovibrational
spectrum.
In all approximations we used the basis set containing 10000 harmonic functions,
i.e. 10 basis set functions for each vibrational degree of freedom, which was enough for the convergence
of at least the 200 lowest eigenvalues.
The Hamiltonian matrix consists of the sub-blocks pertaining to each $K=-J,J$.
However, these subblocks are not independent, since they are connected through the dependence of $\mu_{\alpha \beta}$
on the coordinates $Q$, and the non-diagonal terms become responsible for the rovibrational interactions.
The Davidson algorithm \cite{davidson1975} was used to find the lowest eigenvalues and
the corresponding eigenvectors of the rovibrational Hamiltonian.
The rovibrational energy levels were then assigned by examining the largest weight of the harmonic
part for the vibrational levels and the largest weight of the rigid-rotator eigenfunction.
In this manual  procedure we were able to
localize the first 30 vibrational levels. An unambiguous identification for higher energy values turned out
to be impossible
because of the existence of
anharmonic eigenfunctions almost equally distributed between some harmonic
eigenfunctions.
Special care has to be taken to ensure proper description of the vibration motion within the C--H bond.  Because of the presence of the light hydrogen atom this mode is floppy and highly anharmonic.  In order to improve this description the
``extended''
harmonic basis functions have been added to the basis for
for the $\nu_{1}$-mode, corresponding to the C--H stretch, utilizing
the scaled exponents $\alpha k_{s}$
instead of the harmonic force constants $k_{s}$ in \eref{eq:phil},
where $\alpha$ was chosen from the interval (0.5--1.0) to enable better convergence.

\section{Results and discussion\label{s6}}
In Table \ref{tab0} equilibrium distances calculated in this work for the molecules under study
are presented and compared with the experimental values.
An inspection of this Table shows that for all molecules the difference between the experimental and the {\em ab initio} distances between the carbon atom with the hydrogen atom is small, about 0.002~\AA.
Somewhat larger differences, 0.007 and 0.008~\AA, are observed for the second distance if the third-row atom is connected to the carbon atom. Most probably, this is a result of neglecting the correlation of
the core electrons. However, as pointed out in Ref. \cite{Martin:1995}, such a small discrepancy has only a minor effect on the calculated rovibrational constants. Therefore we decided to proceed with the calculations of the
CCSD(T) force constants, \eref{eq:vq}, with only valence electrons correlated.

For the HCN molecule the importance of various terms in Eqs. (\ref{eq:taylor})--(\ref{eq:hvib4})
on the computed  vibrational energy levels have been examined.
The results of these test calculations are listed in Table \ref{tab:hcn_vib_2}.
First, we note that
including all the potential energy terms and keeping the simple harmonic oscillator
expression for the kinetic energy, cf. the column ``Harm.+$V$'',
leads to an imbalance in the included contributions
which results in an overestimation of the energy levels and in somewhat worse results than the simple harmonic
approximation.
Adding the anharmonic kinematic coefficients $\chi_{spr}$  and $\chi_{sprt}$ from \eref{eq:chi}
which, together with the potential energy function expanded up to and including the fifth order,
lowers the energies making them closer to the experiment. Finally, it is interesting to examine
the effect of adding the fifth-order terms $\chi_{sprtu}$ in \eref{eq:taylor}.
It turns out that within the latter approximation the errors are reduced only by 1-2~cm$^{-1}$ for the lowest energy levels. Therefore,
we decided to skip $\chi_{sprtu}$ in the production calculations. Including of the fifth-order
kinematic coefficients might be necessary for higher vibrational energies, especially in view of the fact that
the absolute error for the series $0\nu_2 0$ increases nonlinearly with $\nu_2$.
However, when dealing with the expansion of the kinetic and potential terms, it is important to take all values of the same order in the Taylor expansion series,  because  neglecting even some of them may lead to high discrepancies in the final results. This follows from the fact that the variational procedure is very dependent on the non-diagonal terms included in the Hamiltonian matrix.
Finally, it is interesting to notice that Approximation~2a gives worse results than Approximation~2 for two high-$l$ cases
presented in the Table, while for all other cases the latter approach gives slightly worse agreement with the experiment.

For linear molecules the bending coordinates $\{Q_{2a},Q_{2b}\}$ are present only as even powers in
the Taylor expansions,
 so $V(Q_{2a},Q_{2b})$ is even with respect to $Q_{2a}$ and $Q_{2b}$.
This fact is due to the symmetry of the degenerate bending mode with respect to
the plane perpendicular to the plane of vibration.
We can also note an interesting property of high order force constants containing the bending coordinates,
coming from the equivalence of the two bending vibration planes:
\begin{align*}
&
F_{2_a,2_a,2_a,2_a}=F_{2_b,2_b,2_b,2_b}=F_{2_a,2_a,2_b,2_b}/3.
\end{align*}

\begin{figure}[h]
\caption{Comparison of the present method with the experiment for the bending mode of HCN.
 \label{fig2}
}
 \begin{center}
 \includegraphics[width=15.0cm,angle=0]{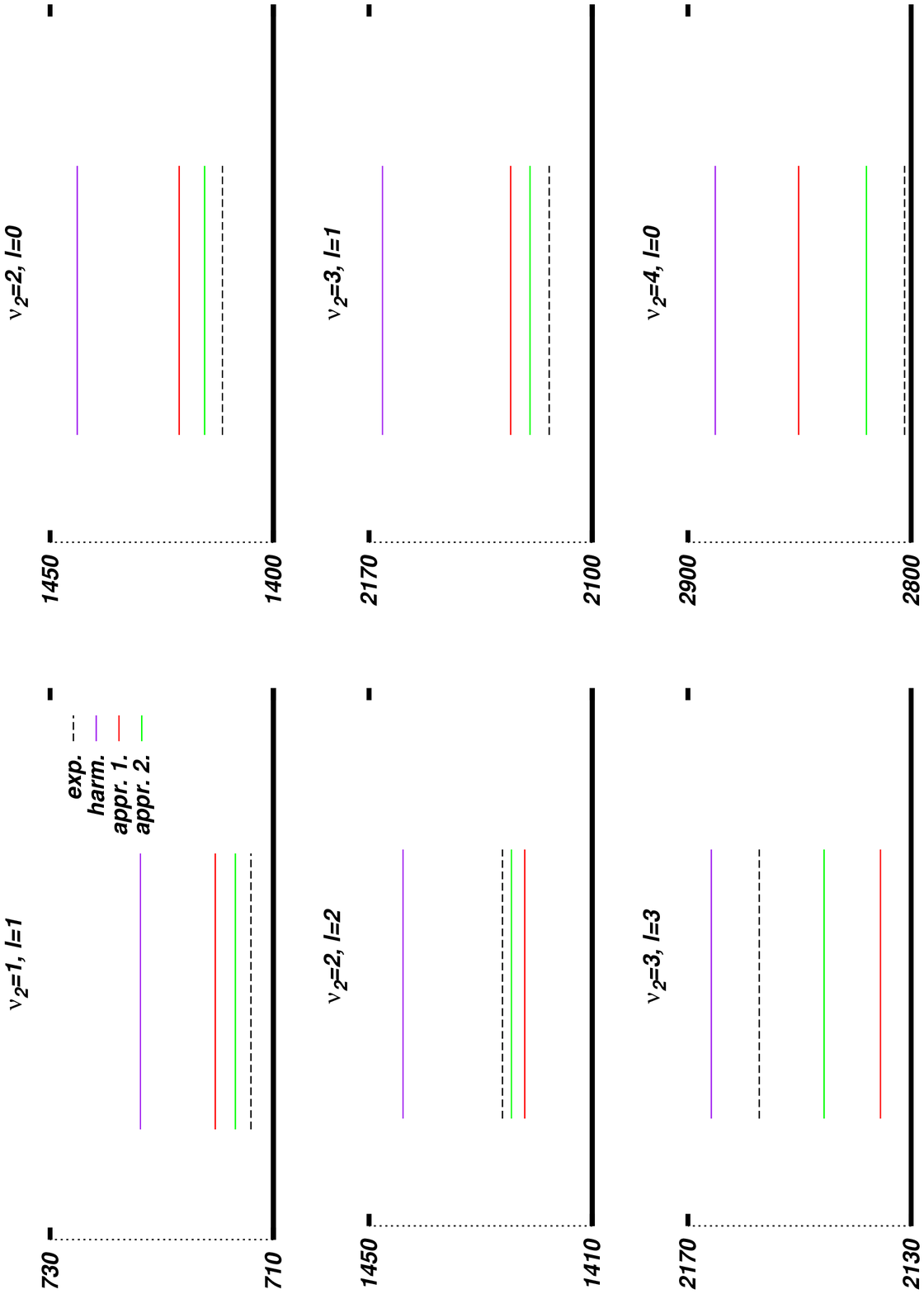}
 \end{center}
\end{figure}

\begin{figure}[h]
\caption{Comparison of the present method with the experiment for the bending mode of HCP.
 \label{fig3}
}
 \begin{center}
 \includegraphics[width=15.0cm,angle=0]{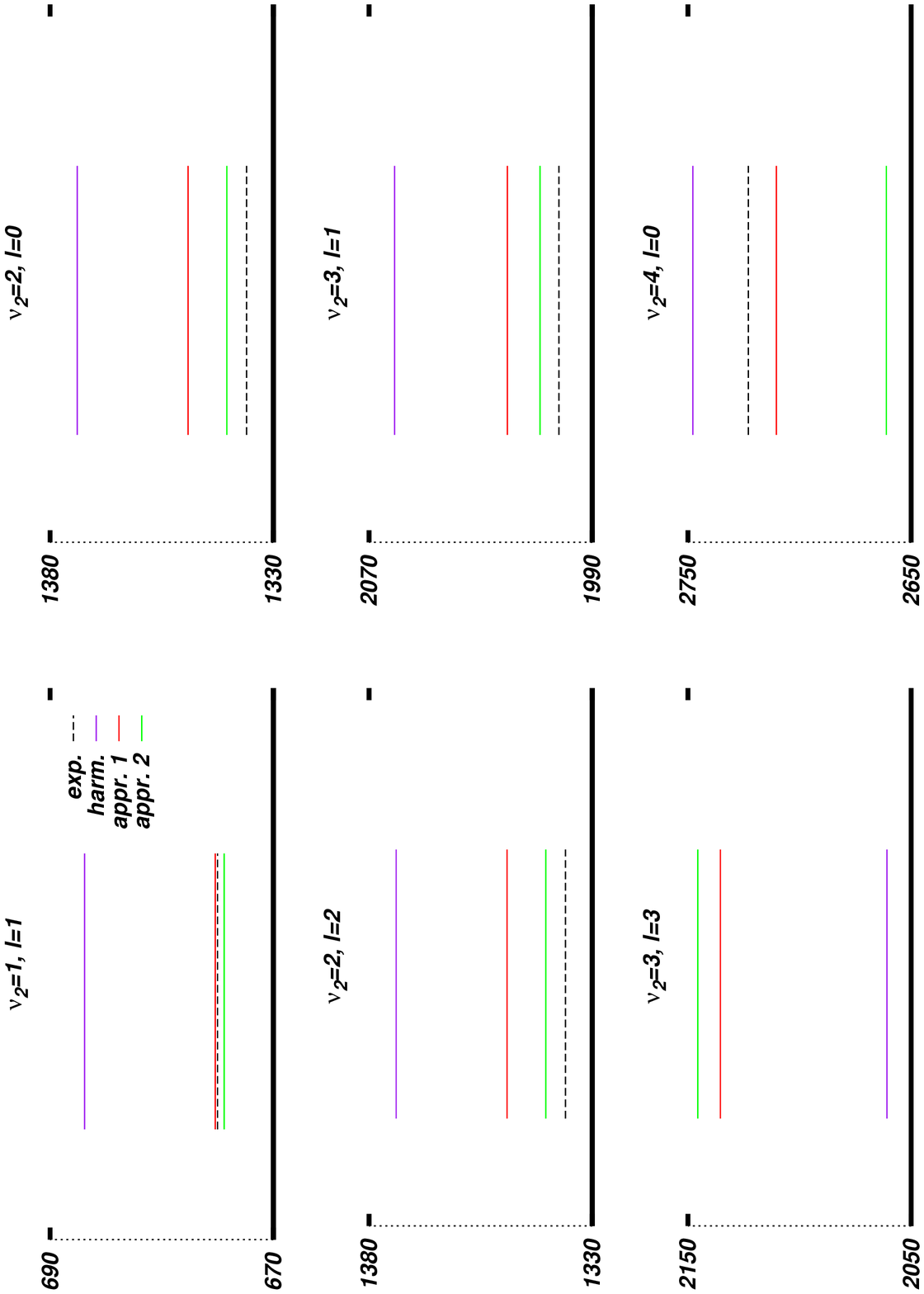}
 \end{center}
\end{figure}

\begin{figure}[h]
\caption{Comparison of the present method with the experiment for the bending mode of OCS.
 \label{fig4}
}
 \begin{center}
 \includegraphics[width=15.0cm,angle=0]{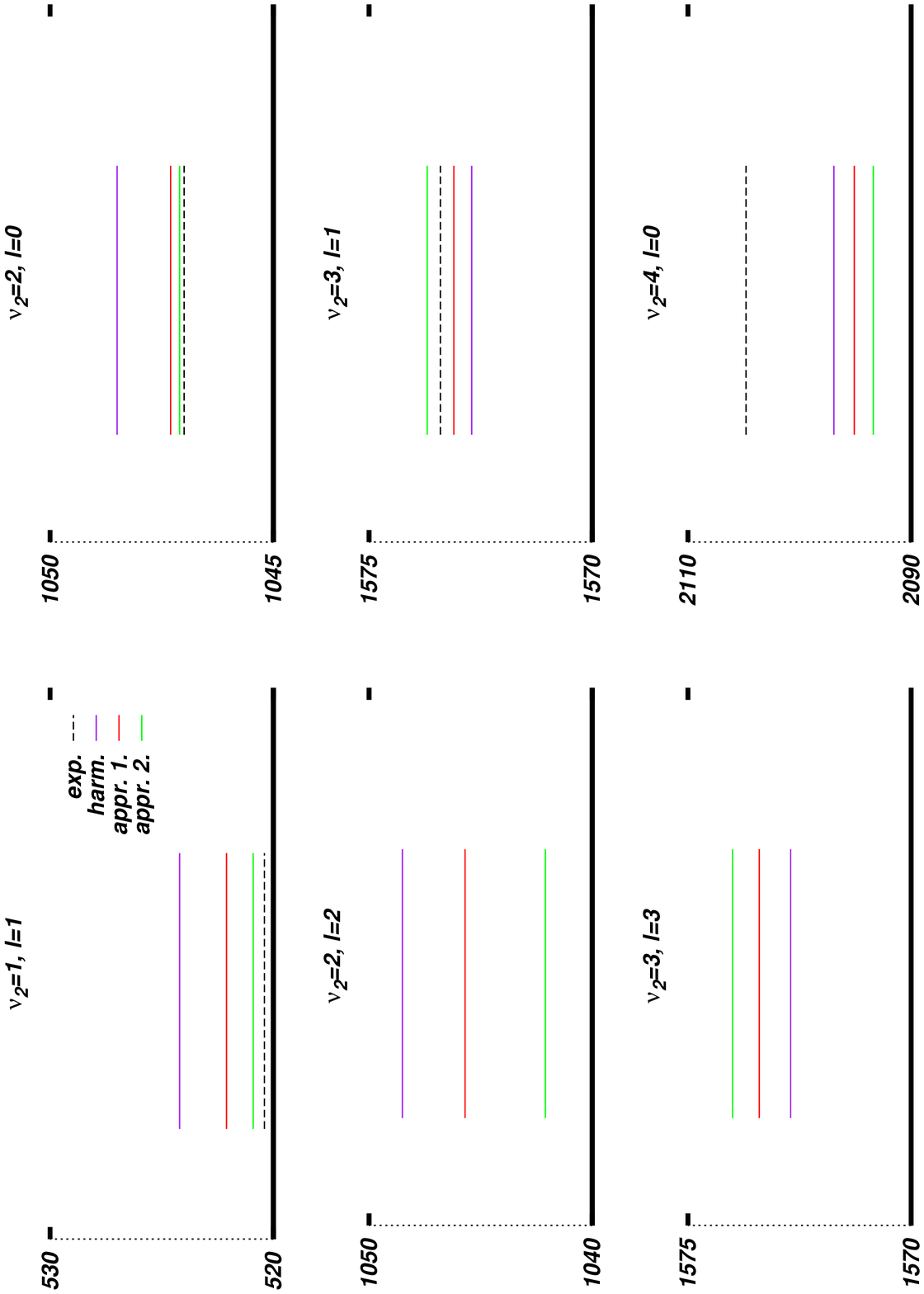}
 \end{center}
\end{figure}

The pure vibrational energy levels relative to the ground-state energy level are listed in Tables \ref{tab:hcn}--\ref{tab:hcp}
for the molecules HCN, OCS, and HCP, respectively.
In these tables the fundamental energy levels as well as some overtones are presented.
Three approximations have been used in these tables: harmonic (Approximation~0), and Approximations~1 and~2 differing
in the way in which the $V$ operator was truncated, cf. Section \ref{sec:tech}.
The results are compared with the experimental data when available.
The experimental data for the energies and rotational constants were taken from Ref. \cite{mellau2008} for HCN,
Refs. \cite{fayt1972,schneider1989,hornberger1996} for OCS, and Refs. \cite{lehman1985a,jung1997} for HCP.
Additionally, Figures \ref{fig2}--\ref{fig4} show selected energy levels for lowest few bending modes of HCN,
HCP, and OCS calculated via all  three methods and compared with the experiment.
Differences between the theoretical and the experimental results are also reported for better readability of the tables.

The results reported in Tables \ref{tab:hcn}--\ref{tab:hcp}
show that in the series of approximations, starting with the harmonic and ending
with Approximation~2, an increasing accuracy is observed in nearly all cases.
Obviously, the harmonic approximation gives the largest errors with respect to the experimental values, and
 Approximation~1 always improves the agreement with the experiment.
The level of improvement varies substantially depending on the molecule and the vibrational mode studied,
but often the use of Approximation~1 causes even a five-fold lowering
of the energy gap between the calculated and experimental energy levels.
Approximation~2 leads in almost all cases to results even closer to the experiment than Approximation~1.
It is also interesting to note that our calculations support the claim concerning the high anharmonicity of the C--H stretching
mode. For the HCN and HCP molecules these modes are easily identified by looking at the
$\Delta_0$ column. They are the only states with errors larger than 100~cm$^{-1}$.
Although both Approximations 1 and especially 2 do a good job of bringing this mode close to the experiment,
it could be envisaged that the next approximation would give an even better agreement.
Unfortunately,
the analytical gradients within the CCSD(T) approach should be used for this purpose. Another way to improve
the potential energy surface is to use Morse-like potential for this mode, which
corresponds approximately
to the C--H stretch. Such potential better describes upper vibrational states close
to the dissociation limit, see for example Ref. \cite{bowman1993}. Unfortunately,
this leads to serious complications in the calculations of the matrix elements
and has not been attempted in this paper.

In Tables \ref{tab:hcn}--\ref{tab:hcp} a comparison of
our results with previous calculations for every molecule is also presented.
As already mentioned above,
HCN is the most studied molecule and a vast theoretical material is available in the literature
dealing with the problem of the vibrational HCN spectra.
In Table \ref{tab:hcn} we listed values from four different
sources.
The first column with the theoretical results corresponds to the earliest work utilizing an \textit{ab initio}  PES \cite{bowman1993},
where one of the modifications of the DVR method was used to calculate the vibrational energies.
The next set of results
stems from
a more recent implementation of the DVR method
\cite{tennyson2001}.
It should be stressed again that although the DVR method allows finding of vibrational energies up to very high excitation
levels, it is limited to triatomic molecules and is therefore less general than the method presented in this paper in the curvilinear
coordinates.
Another work applying the DVR approach is Ref. \cite{csaszar2007}, where DVR was implemented for the Watson Hamiltonian \cite{wf1,wf2,wf3}.
The last set of data was produced by Wang {\em et al.} \cite{wang2000} by applying a finite element method
for the Watson Hamiltonian with $J=0$.
A perusal of this part of Table \ref{tab:hcn} allows us to conclude that the present vibrational energies
are generally better than those of Ref. \cite{bowman1993} and of the same level of accuracy than in the other
references, except DVR implementation made in \cite{tennyson2001} where the accuracy is better,
especially for combination bands, like (2 0$^0$ 3) in Table \ref{tab:hcn}.  
This fact is reassuring in view of the just mentioned possible high-precision of the DVR method
for the triatomic molecules.

For carbonyl sulfide much fewer theoretical investigations are available.
Since the OCS molecule is rigid, the model
with the neglected
bending mode can be used to study vibrationally excited stretching states.
Such an approach has been pursued by Peterson {\em et al.}
\cite{Peterson:1991}, The vibrational energies obtained in Ref. \cite{Peterson:1991} are listed in Table \ref{tab:ocs}.
It is not surprising that the discrepancy between these results with the experimental values
is larger than in the case of our approach, where
the mixed bending-stretching terms have been taken into account. To our knowledge it was the only
\textit{ab initio} available potential energy surface for excited vibrational states.
The next column of Table \ref{tab:ocs} contains the data of Ref. \cite{sarkar2009},
where an algebraic model was used to analyze and interpret
 the experimental rovibrational spectra of small and
medium-sized molecules. This method employs the Lie algebra techniques to obtain an effective Hamiltonian
operator which describes rovibrational degrees of freedom. Since it does not
use any \textit{ab initio} results, but instead is solely based on the fitting of the spectroscopic data, the results obtained there
are very close to the experimental values. Finally, the last
theoretical column in  Table \ref{tab:ocs} contains
data from Ref. \cite{aubanel1988}, where a semi-classical model based on the adiabatic switching method (ASM)
was used together with a spectroscopic potential energy function.
Also, in this case the good agreement with the experiment can be attributed to the use of some spectroscopic data
in the actual calculations.
Taking this into account we can conclude that our method reproduces the vibrational energies for OCS
much better that the other {\em ab initio} approach \cite{Peterson:1991}.

The DVR method was also applied
to the phosphaethyne molecule \cite{beck2000}.
The solution to the
vibrational problem with an \textit{ab initio} PES gave
,in general, a slightly better agreement with the experiment than our results.
The following  column in Table \ref{tab:ocs} shows the results of Ref. \cite{koput1997}, where an \textit{ab initio} PES was used to computed
the rovibrational energy levels with a  mixed variational-pertubation theory approach. Also here
a somewhat better agreement between the vibrational energy levels and the experiment is found.
However, the constants of the rovibrational interactions were calculated only 
for the ground state in this paper.

Several possible reasons of the remaining discrepancies should be considered.
The first issue is the quality of the CCSD(T)/cc-pVQZ electronic energies.
Since the CCSD(T) method is known to be very accurate for molecules (unless very large distortions
to the geometry are studied) and the cc-pVQZ orbital basis is also quite accurate, we can
exclude the possibility that inaccurate electronic energies are the reason for the remaining errors.
The most probable cause of the discrepancies are the truncation of the Taylor expansions which we have used for the $g$-tensor and for the potential $V$ in the Hamiltonian.
For instance, we used the second-order (i.e. linear and quadratic)
expansion in the vibrational coordinates of $\mu_{\alpha \beta}(Q)$, which may be insufficient
for some excited rovibration bending energy levels, if the average geometry of the molecule
differs considerably from the equilibrium geometry.

For the C$\equiv$N, C$\equiv$P, and C=S stretching modes, generally larger errors
are observed for the
fundamental transition than for the other two. Errors of about 5~cm$^{-1}$ can be seen
in the former case and only 1-2~cm$^{-1}$ for the latter.
The error increases approximately twofold for the first overtone ($20^00$).
According to Koput {\em et al.} \cite{koput1997} this discrepancy can be explained
by the effect of neglecting the correlation between the core electrons
of the heavier atoms. This hypothesis is supported by the fact that only the $\nu_1$ mode
is so atrongly affected. It seems that further studies of the molecules containing atoms from
the third row of the periodic table should account for the core correlation.
Similar discrepancies have been found by Martin {\em et al.} \cite{Martin:1995}.

Finally, one more complication arises from the fact that
the values of the matrix elements $\mb$ are very sensitive to small non-diagonal
anharmonic elements of the potential function $V$. These elements have a small influence
on the calculated rovibrational energy levels, but cause a considerable change in
the wave functions, and, hence in the average geometry of the molecule and its rotational
energy levels if the molecule is vibrationally excited.

In Tables \ref{hcn_rot_1}--\ref{hcp_rot_1} the constants for the rovibrational energies calculated
within Approximation~2 are presented. The experimental values for these molecules are also given, when available.
These constants have been obtained by first
computing rovibrational energy levels for various values of $J$ and $K$, and then by performing fitting of the $E_r$ values to the analytic form, cf. Eqs. (\ref{eq:er})--(\ref{eq:er0}). The $J$ values up to and including $J$=15 were used.
The results show that the most important constants,
i.e. $B_v$, $D_v$, and $q_v$,
are reproduced quite accurately. For the higher-order constants, the accuracy
of the calculated rovibrational energies and the truncation of the maximum $J$ was found to be
insufficient, so the
$H_v$ and $q_{vJ}$ constants are not reported in Tables \ref{hcn_rot_1}--\ref{hcp_rot_1}.
Apart from the comparison to the experiment it
is interesting to examine the dependence of the rotational constants on the vibrational quantum numbers.
Such a dependence is especially visible for molecules containing a hydrogen atom,
like HCN and HCP.
The results allow us to conclude that
the constants $B_{\nu}$ and $D_{\nu}$ grow linearly with $\nu_{2}$,
the quantum number for the degenerate bending mode, while for fixed $\nu_2$
they decrease with $|l|$, the vibrational angular momentum quantum number for the bending mode $\nu_2$.
This finding can be explained by the fact that for a molecule in an excited bending mode,
the component of $I_{z}$ becomes smaller which leads to a lowering of the constants $B_{\nu}$ and $D_{\nu}$.
On the other hand,
decreasing the constants while increasing $|l|$ is mostly caused by the interaction of the total angular momentum with the vibrational one.
These constants also become smaller when the quantum numbers $\nu_1$ and $\nu_3$ increase.
This result can be rationalized by the fact as the vibrationally
averaged interatomic distance $\langle r\rangle$ becomes larger, the distance differs more from the equilibrium distance.
This assumption can be supported by calculating the vibrationally averaged values of
the bond distances fo the molecules under study. Luckily, such calculations were reported by
Laurie {\em et al.}  \cite{laurie1962} who presented
$\langle r_{\text{HC}} \rangle =1.0739 \AA$, $\langle r_{\text{CN}} \rangle =1.1574$ \AA\ for the ground state.
These values are larger than the equilibrium values from Table \ref{tab0}.

\section{Summary\label{s7}}
The rovibrational Hamiltonian in curvilinear vibrational coordinates has been used to
 solve the nuclear motion problem for several linear triatomic molecules in curvilinear
vibrational coordinates. The curvilinear coordinates were shown to have advantages
over the more commonly used linear coordinates. Comparison with the exising theoretical
results shows that the present approach works equally well comparing to other {\em ab initio} calculations
of vibrational energy levels, and is not limited to triatomic molecules.
In its current form the approach presented in this paper provides a good accuracy
for most vibrational levels and recovers the rotational and the $l$-doubling constants for the studied molecules.
Further improvements to the code will also include the calculation of  intensities, so that the
full spectrum of the molecules can be simulated.

\section*{Acknowledgments}

This work was supported by the Polish Ministry of Science and
Education through  project N N204 215539.

\clearpage
\newpage

\clearpage
\newpage


\begin{table}[h!]
\caption{Equilibrium bond lengths (in \AA) obtained in this work and taken from the experiment.
$r_1$ ($r_2$) denotes the distance between carbon and a lighter (heavier) atom.
\label{tab0}
}
\begin{center}
\begin{tabular}{lllll}
\hline
Molecule&  $r_{1}$&$r_{2}$ & Reference\\
\hline
HCN &  1.06686 & 1.15646 & this work \\
    &  1.065   & 1.153   & \cite{Winnewisser:1976} \\
HCP &  1.07216 & 1.54732 & this work \\
    &  1.070   & 1.540 & \cite{Drean:1996} \\
OCS &  1.15830 & 1.56902 & this work \\
    &  1.156   & 1.561   & \cite{Lahaye:1987} \\
\hline
\end{tabular}
\end{center}
\end{table}


\begin{table}[h!]
\begin{center}
\caption{The influence of the kinematic coefficients on the vibrational energies for the HCN molecule.
All values in cm$^{-1}$. Approximation~2a with added $\chi_{sprtu}$
\label{tab:hcn_vib_2}
}
\end{center}
\begin{center}
\begin{tabular}{rrrrrr}
\Xhline{1.2pt}
$\nu_{1}\nu_{2}^{l}\nu_{3}$ & Harm. & Harm. + $V$ &Appr. 2&Appr. 2a&Exp.\\
\hline
0 1$^1$ 0 &721.9&727.6&713.4&712.9&712.0 \\
0 2$^0$ 0 &1443.9&1447.9&1415.4&  1413.7&    1411.4  \\
0 2$^2$ 0 &1443.9&1460.4&1424.5& 1423.3 & 1426.1  \\
0 3$^1$ 0 &2165.8&2170.4&2119.6&2117.7 & 2113.5   \\
0 3$^3$ 0 &2165.8&2190.8&2145.6& 2143.3& 2157.2  \\
0 4$^0$ 0 &2887.8&2926.5&2820.1& 2816.1 &  2803.0  \\
0 0$^0$ 1 &2123.0&2148.5& 2098.1& 2098.0&   2096.4  \\
1 0$^0$ 0 &3435.3&3554.5&3316.8&3314.6 &3311.4  \\
0 0$^0$ 2 &4246.0&4231.5& 4168.2&  4165.3 &  4173.2  \\
2 0$^0$ 0 &6870.7&6934.7&6529.4& 6524.9& 6519.5   \\
0 1$^1$ 1 &2844.9&2898.3&2810.5& 2807.4 & 2805.6  \\
2 0$^0$ 3 &13239.6&13432.9&12675.1&12669.2& 12658.0 \\
\hline
\end{tabular}
\end{center}
\end{table}


\begin{table}[h]
\begin{center}
\caption{Experimental and theoretical vibrational energy levels
for the HCN molecule.
Theoretical results were obtained within the harmonic approximation (Approximation~0),
and with  Approximation~1 and Approximation~2 defined in the text.
$\Delta_i$ denotes the error of a given approximation $i$ with respect to the experiment, i.e.
$\Delta_i=E(\mathrm{Appr.~i})-E(\mathrm{Exp})$.
All values are in cm$^{-1}$.
Experimental data are taken from Ref. Ref. \cite{mellau2008}.
\label{tab:hcn}
}
\end{center}
\begin{center}
\resizebox{10cm}{!} {
\begin{tabular}{rrrrrrrrrrrr}
\Xhline{1.2pt}
$\nu_{1}\nu_{2}^{l}\nu_{3}$&Harm.& $\Delta_{0}$&Appr. 1&$\Delta_{1}$&Appr. 2&$\Delta_{2}$&Exp
& Ref. \cite{bowman1993} & Ref. \cite{tennyson2001} & Ref. \cite{csaszar2007} & Ref.  \cite{wang2000}\\
\hline
0 1$^1$ 0 &721.9&9.9&715.2&3.2&713.4&1.4&712.0 &718.4   &712.0 &715.9    &713.0\\
0 2$^0$ 0 &1443.9&32.5&1421.1&  9.7&    1415.4 &4.0&    1411.4 &1418.9  &1411.4 &1414.9    &1421.6\\
0 2$^2$ 0 &1443.9&17.8&1422.1& -4 & 1424.5    &-1.6&1426.1&n/a     &n/a     &n/a        &n/a\\
0 3$^1$ 0 &2165.8&52.3&2125.6& 12.1 & 2119.6    &6.1&2113.5&2126.0  &2113.45 &n/a          &2127.9\\
0 3$^3$ 0 &2165.8&21.8 &2135.5&  -8.5 &  2145.6   &1.6&2157.2&n/a     &n/a     &n/a        &n/a\\
0 4$^0$ 0 &2887.8&85.8&2850.4& 37.4  &  2820.1    &18.1&2803.0&2812.5  &2807.1 &2801.5    &2834.3\\
0 0$^0$ 1 &2123.0&26.6&2095.1& -1.3 &   2098.1  &1.7&2096.4&2090.3  &2091.0 &2100.6    &2083.2\\
1 0$^0$ 0 &3435.3&123.9&3313.4& 2 &  3316.8   &5.4&3311.4&3334.1  &3311.5 &3307.7    &3340.4\\
0 0$^0$ 2 &4246.0&72.8&4181.7& 8.5  &  4168.2  &-5&4173.2&4161.5  &4173.1 &4181.5    &4146.3\\
2 0$^0$ 0 &6870.7&351.2&6598.4& 78.9  & 6529.4   &-9.9&6519.5 &6553.2  &6519.6 &6513.5    &n/a \\
0 1$^1$ 1 &2844.9&39.3&2809.4& 3.8  & 2810.5   &4.9&2805.6 &2806.4  &2807.1 &n/a        &n/a\\
2 0$^0$ 3 &13239.6&581.6&12795.1&137.1 & 12675.1  &17.1&12658.0 &n/a     &12658.0 &n/a        &n/a\\
0 2$^0$ 1 &3566.9&64.8&3532.1&30.0 & 3512.0  &9.9&3502.1 & 3507.4 & 3511.0 & 3511.0 & 3511.0 \\
0 2$^2$ 1 &3566.9&44.2&3541.7&19.0 & 3529.4  &6.7&3522.7 & n/a & n/a & n/a & n/a \\
1 3$^1$ 0 &5601.1&232.8&5435.7&67.4 & 5398.4  &30.1&5368.3 & 5387.1 & n/a & n/a & n/a \\\\
\hline
\end{tabular}
}
\end{center}
\end{table}

\begin{table}[h]
\begin{center}
\caption{Experimental and theoretical vibrational energy levels for the OCS molecule.
All values are in cm$^{-1}$.
See Table \ref{tab:hcn} for an explanation of the symbols.
Experimental data are taken from Ref. \cite{fayt1972}.
\label{tab:ocs}
}
\end{center}
\begin{center}
\resizebox{10cm}{!} {
\begin{tabular}{rrrrrrrrrrr}
\Xhline{1.2pt}
$\nu_{1}\nu_{2}^{l}\nu_{3}$&Harm.& $\Delta_{1}$&Appr. 1&$\Delta_{2}$&Appr. 2&$\Delta_{3}$&Exp
& Ref.  \cite{Peterson:1991} & Ref. \cite{sarkar2009} & Ref. \cite{aubanel1988}\\
\hline
0 1$^1$ 0 &524.2&3.8&522.1&1.7&520.9&0.5&520.4 &n/a&n/a          &520.3\\
0 2$^0$ 0 &1048.4&1.4&1047.3&  0.3&    1047.1 &0.1&1047.0&n/a&1047.0    &1046.9\\
0 2$^2$ 0 &1048.4& &1045.7&   & 1042.1& &n/a&n/a&n/a          &1048.8\\
0 3$^1$ 0 &1572.7&-0.7&1573.1&-0.3 & 1573.7    &0.3&1573.2&n/a&1569.0    &1573.4 \\
0 3$^3$ 0 &1572.7& &1573.4& & 1574.0   & &n/a&n/a&n/a          &1561.5\\
0 4$^0$ 0 &2096.9&-7.9&2101.3& -3.5  &  2102.4    &-2.4&2104.8&n/a&2088.0    &2105.8\\
0 0$^0$ 1 &871.7&12.4&861.4& 2.1 &   859.1  &-0.2&859.3&869.0&855.5     &859.2\\
1 0$^0$ 0 &2094.7&32.4&2078.4&16.2  &  2067.1   &4.9&2062.2&2071&2057.9    &2062.1\\
0 0$^0$ 2 &1743.4&32.7&1715.7&  5 &  1712.7  &2&1710.7&1730.0&1705.0    &1710.6 \\
2 0$^0$ 0 &4189.3&86.9&4127.2& 25.8  & 4110.7   &9.3&4101.4&4119.0&4097.1    &4101.4\\
0 1$^1$ 1 &1395.9& &1379.4&   & 1373.1   &n/a&n/a &n/a&n/a          &1372.8\\
2 0$^0$ 3 &6804.2& &6704.1&  & 6687.5  & &n/a&n/a&n/a          &n/a\\
0 2$^0$ 1 &1920.1&27.9&1915.2&23.0 & 1899.7  &7.5&1892.2&n/a&1895.5       &1891.7\\
0 2$^2$ 1 &1920.1& &1911.4& & 1890.5  & &n/a&n/a          &1886.7      &1890.5\\
1 3$^1$ 0 &3667.4&52.0&3643.2&27.8 & 3618.8  &3.4&3615.4&n/a&3617.9       &3614.4\\
\hline
\end{tabular}
}
\end{center}
\end{table}

\begin{table}[h]
\begin{center}
\caption{Experimental and theoretical vibrational energy levels
for the HCP molecule.
All values are in cm$^{-1}$.
See Table \ref{tab:hcn} for an explanation of the symbols.
Experimental data are taken from Refs. \cite{lehman1985a}$^{a}$ and \cite{jung1997}$^{b}$.
\label{tab:hcp}
}
\end{center}
\begin{center}
\resizebox{10cm}{!} {
\begin{tabular}{rrrrrrrrrrr}
\Xhline{1.2pt}
$\nu_{1}\nu_{2}^{l}\nu_{3}$&Harm.& $\Delta_{0}$&Appr. 1&$\Delta_{1}$&Appr. 2&$\Delta_{2}$&Exp
& Ref. \cite{beck2000} & Ref. \cite{koput1997} & Ref. \cite{sarkar2009}\\
\hline
0 1$^1$ 0 \hspace{1em}$^{a}$ &686.9&11.95&675.2&0.2&674.4&-0.6&675&675    &675.2   &n/a \\
0 2$^0$ 0 \hspace{1em}$^{a}$ &1373.9&37.9&1349.1&  13.1&    1340.4 &4.4&1336&1336  &1335.0  &1332.9\\
0 2$^2$ 0 \hspace{1em}$^{a}$ &1373.9&9.9&1369.1& 5.1 & 1365.7    &1.7&1364&1364   &n/a      &1340.2\\
0 3$^1$ 0 \hspace{1em}$^{a}$ &2060.8&58.8&2020.4&18.4 & 2008.7    &6.7&2002 &2002   &2002.1  &1995.6\\
0 3$^3$ 0 \hspace{1em}$^{a}$ &2060.8& &2041.1&   & 2036.1   & &n/a&n/a       &n/a      &n/a\\
0 4$^0$ 0 \hspace{1em}$^{b}$ &2747.8&105.0&2721.1& 78.3 &  2715.8   &73.0&2643&n/a       &2653.8  &n/a\\
0 0$^0$ 1 \hspace{1em}$^{a}$ &1294.9&15.9&1287.2&8.2&   1285.1 &6.1&1279&n/a       &1275.5   &1279.0\\
1 0$^0$ 0 \hspace{1em}$^{b}$ &3345.3&129.3&3301.7&85.7 & 3207.1   &-8.9&3217&n/a       &3215.3   &n/a\\
0 0$^0$ 2 \hspace{1em}$^{a}$ &2589.8&40.8&2571.4& 22.4 &  2570.1  &21.1&2549&n/a       &2540.1  &2547.4\\
2 0$^0$ 0 \hspace{1em}$^{a}$ &6690.5& &6671.4&   & 6667.8   & &n/a&n/a       &6321.0  &n/a\\
0 1$^1$ 1 \hspace{1em}$^{a}$ &1981.8&34.8&1960.1& 13.1  & 1955.1   &8.1&1947&n/a       &n/a      &n/a \\
2 0$^0$ 3 \hspace{1em}$^{a}$ &10575.2& &10384.7&  & 10276.7  & &n/a&n/a       &n/a      &n/a\\
0 2$^0$ 1 \hspace{1em}$^{a}$ &2668.8&70.8&2634.2&36.2& 2608.4 &10.4&2598 &2598       &n/a      &2597.2\\
0 2$^2$ 1 \hspace{1em}$^{a}$ &2668.8&42.8&2647.2&21.2&2636.3&10.3&2626&2626       &n/a      &2604.5\\
1 3$^1$ 0 \hspace{1em}$^{a}$ &5406.1& &5302.7& & 5249.7 & &n/a&n/a        &n/a      &n/a\\
\hline
\end{tabular}
}
\end{center}
\end{table}

\begin{table}[h!]
\begin{center}
\caption{Rovibrational constants for
 HCN. Theoretical results were obtained with Approximation~2.
All values are in cm$^{-1}$.
Experimental data are taken from Ref. \cite{mellau2008}.
\label{hcn_rot_1}
}
\end{center}
\begin{center}
\begin{tabular}{rrrrrrr}
\Xhline{1.2pt}
$\nu_{1}\nu_{2}^{l}\nu_{3}$ &$B_{v}$ Exp  &$B_{v}$ Calc &$D_{v}\times 10^{6}$ Exp&$D_{v}\times 10^{6}$ Calc&$q_{v}\times 10^{3}$ Exp& $q_{v}\times 10^{3}$ Calc\\

\hline
0 0$^0$ 0 &1.478222&1.479782 &2.91028&2.59259 &&\\
0 1$^1$ 0 &1.481773&1.481931 &2.97746&3.01327 &7.48773  &7.51431\\
0 2$^0$ 0 &1.485828&1.487207 &3.04799&3.08957 &7.59561  &7.63292\\
0 2$^2$ 0 &1.484997&1.483982 &3.04079&3.11047 &       & \\
0 3$^1$ 0 &1.489575&1.492028 &3.11365&3.19783 &7.70926  &7.81492\\
0 3$^3$ 0 &1.487868&1.491873 &3.10024&3.29147 &       & \\
0 1$^1$ 1 &1.471574&1.475982 &2.98254&3.11873 &7.47954  &7.49425\\
0 2$^0$ 1 &1.475493&1.482542 &3.05017&3.78874 & 7.57341 &7.61422\\
0 2$^2$ 1 &1.474678&1.491012 &3.04520&3.85578 &       & \\
1 3$^1$ 0 &1.479809&1.481287 &3.11952&4.50034 &7.85215  &7.83421\\
\Xhline{1.0pt}
\end{tabular}
\end{center}
\end{table}

\begin{table}[h!]
\begin{center}
\caption{Rovibrational constants for
 OCS. Theoretical results were obtained with Approximation~2.
All values are in cm$^{-1}$.
Experimental data are taken from Refs. \cite{schneider1989}$^{a}$ and
\cite{hornberger1996}$^{b}$.
\label{ocs_rot_1}
}
\end{center}
\begin{center}
\begin{tabular}{lrrrrrr}
\Xhline{1.2pt}
$\nu_{1}\nu_{2}^{l}\nu_{3}$
&$B_{v}$ Exp &$B_{v}$ Calc&$D_{v}\times 10^{8}$ Exp&$D_{v}\times 10^{8}$ Calc&$q_{v}\times 10^{4}$ Exp& $q_{v}\times 10^{4}$ Calc\\
\hline
0 0$^0$ 0 \hspace{1em}$^{a}$ &0.202856  &0.202506  &4.341064  &4.703412   & &                \\
0 1$^1$ 0 \hspace{1em}$^{a}$ &0.203209  &0.203347  &4.411467  &5.198434   &2.12193 &2.11932  \\
0 2$^0$ 0 \hspace{1em}$^{b}$ &0.203480  &0.203135  &3.66256   &3.963137   &n/a     &2.13421  \\
0 2$^2$ 0 \hspace{1em}$^{b}$ &0.203559  &0.203397  &5.23854   &4.598443   &      &   \\
0 3$^1$ 0                    &n/a       &0.203492  &n/a       &4.993237   &n/a     &2.16402  \\
0 3$^3$ 0                    &n/a       &0.204431  &n/a       &5.194394   &      &   \\
0 1$^1$ 1 \hspace{1em}$^{a}$ &0.202657  &0.203054  &4.54206   &5.94217    &2.28496 &2.25421  \\
0 2$^0$ 1 \hspace{1em}$^{a}$ &0.202953  &0.203384  &4.55522   &5.53215    &2.2238  &2.18401  \\
0 2$^2$ 1 \hspace{1em}$^{a}$ &0.203048  &0.203043  &4.63515   &5.73263    &      &   \\
1 3$^1$ 0                    &n/a       &0.203349  &n/a       &6.143694   &n/a     &2.17754  \\
\Xhline{1.0pt}
\end{tabular}
\end{center}
\end{table}

\begin{table}[h!]
\begin{center}
\caption{Rovibrational constants for
 HCP. Theoretical results were obtained with Approximation~2.
All values are in cm$^{-1}$.
Experimental data are taken from Ref. \cite{jung1997}.
\label{hcp_rot_1}
}
\end{center}
\begin{center}
\begin{tabular}{rrrrrll}
\Xhline{1.2pt}
$\nu_{1}\nu_{2}^{l}\nu_{3}$
&$B_{v}$ Exp &$B_{v}$ Calc&$D_{v}\times 10^{6}$ Exp&$D_{v}\times 10^{6}$ Calc&$q_{v}\times 10^{3}$ Exp& $q_{v}\times 10^{3}$ Calc\\
\hline
0 0$^0$ 0    &0.666326 &0.667483       &0.70392&0.93820   &&\\
0 1$^1$ 0    &0.666771 &0.667333       &0.71196&0.99482   &1.63032 &1.65339\\
0 2$^0$ 0    &0.667344 &0.668391       &0.719160&1.03945  &1.65970 &1.68432\\
0 2$^2$ 0    &n/a      &0.663641       &n/a&1.09384       &      & \\
0 3$^1$ 0    &0.667784 &0.668380       &0.72502&1.10939   &1.70887 &1.73283\\
0 3$^3$ 0    &n/a      &0.673452       &n/a&1.03988       &      & \\
0 1$^1$ 1    &0.662932 &0.665482       &0.72052&0.928438  &1.56214 &1.49844\\
0 2$^0$ 1    &n/a      &0.668391       &n/a&0.91938       &n/a     &1.61925\\
0 2$^2$ 1    &n/a      &0.663942       &n/a&1.39492       &      & \\
1 3$^1$ 0    &n/a      &0.653393       &n/a&1.03821       &n/a     &1.66754\\
\Xhline{1.0pt}
\end{tabular}
\end{center}
\end{table}


\clearpage
\newpage
\bibliographystyle{spmpsci1}
\bibliography{rovib,molpro_new,ind_lokal_EOM_clone,sexp_clone,rovib_add,rovib_add2}

\end{document}